\documentclass[prd,twocolumn,showpacs,floatfix,amsmath,amssymb,floatfix]{revtex4}
\usepackage{graphicx,color,dcolumn,booktabs,bm}
\usepackage{longtable,lscape}
\usepackage{txfonts}
\usepackage{overpic}
\usepackage{amssymb}
\usepackage{indentfirst}
\usepackage{feynmf}   
\usepackage{slashed}  
\usepackage{cases}
\usepackage{color}

\graphicspath{{Figures/}} %

\begin{document}

\title{Newly observed $D_J(3000)^{+,0}$ and $D_J^*(3000)^0$ as $2P$ states in $D$ meson family}
\author{Yuan Sun$^{1,2}$}
\email{suny07@lzu.edu.cn}
\author{Xiang Liu$^{1,2}$\footnote{Corresponding author}}\email{xiangliu@lzu.edu.cn}
\author{Takayuki Matsuki$^3$}
\email{matsuki@tokyo-kasei.ac.jp}
\affiliation{$^1$Research Center for Hadron and CSR Physics,
Lanzhou University $\&$ Institute of Modern Physics of CAS,
Lanzhou 730000, China\\
$^2$School of Physical Science and Technology, Lanzhou University,
Lanzhou 730000, China\\
$^3$Tokyo Kasei University, 1-18-1 Kaga, Itabashi, Tokyo 173-8602,
Japan}

\begin{abstract}

In this work, we study the newly observed $D_J(3000)$ and $D_J^*(3000)$ through the analysis of mass spectrum and calculation of the corresponding two-body strong decay behaviors. Our results show that $D_J(3000)$ and $D_J^*(3000)$ are explained as the $2P(1^+)$ and $2^3P_0$ states in the $D$ meson family, respectively, which is supported by the calculated masses of these two states and their decay behaviors. As a byproduct, the decay behaviors of $3^1S_0$, $3^3S_1$, $2D(2^-)$, $2^3D_1$, $2D^\prime(2^-)$, $2^3D_3$,
$2P^\prime(1^+)$, $2^3P_2$, $1F(3^+)$, $1^3F_2$, $1F^\prime(3^+)$, and $1^3F_4$ states are also given, which will be helpful to further experimentally study mixings of these $D$ mesons, too.

\end{abstract}
 \maketitle


Very recently the LHCb Collaboration announced several $D_J$ resonances by studying the $D^+\pi^-$, $D^0\pi^+$ and $D^{*+}\pi^-$ invariant mass spectra, which are obtained from the inclusive processes $pp\to D^+\pi^- X$, $pp\to D^0\pi^+X$ and $pp\to D^{*+}\pi^-X$, respectively, where $X$ denotes a system composed of any collection of charged and neutral particles \cite{Aaij:2013sza}.
$D_J^*(3000)^0$ associated with $D_2^*(2460)^0$ and $D_J^*(2760)^0$ exists in the $D^+\pi^-$
invariant mass spectrum. In addition, their charged partners $D_J^*(3000)^+$, $D_2^*(2460)^+$ and $D_J^*(2760)^+$ were observed in the $D^0\pi^+$ mass invariant spectrum. In the $D^{*+}\pi^-$ mass spectrum,
reported were seven $D_J$ mesons, which are $D_1(2420)^0$, $D_2^*(2460)^0$, $D_J^*(2650)^0$, $D_J^*(2760)^0$, $D_J(2580)^0$, $D_J(2740)^0$ and $D_J(3000)^0$. The measurement from LHCb also indicates that $D_J^*(2650)^0$ and $D_J^*(2760)^0$ are natural states while $D_J(2580)^0$, $D_J(2740)^0$ and $D_J(3000)^0$ are unnatural states \cite{Aaij:2013sza}. Here, the $D$ mesons with
$J^P=0^+,1^-,2^+,...$ and $P=(-1)^J$ are the so-called natural states while those with $J^P=0^-,1^+,2^-,...$ are grouped into unnatural states \cite{Aaij:2013sza}.

Among these atates, $D_J(3000)^{+,0}$ and $D_J^*(3000)^0$ are first observed by experiments
as $D_J$ states around 3 GeV. At the present, experiments measured their resonance parameters as
\begin{eqnarray}
m_{D_J(3000)^0}&=&2971.8\pm8.7\, \mathrm{MeV},\\\Gamma_{D_J(3000)^0}&=&188.1\pm44.8\,\mathrm{MeV},\\
m_{D_J^*(3000)^0}&=&3008.1\pm4.0\,\mathrm{MeV},\\\Gamma_{D_J^*(3000)^0}&=&110.5\pm11.5\,\mathrm{MeV}.
\end{eqnarray}
and they gave the information on several decay channels.
Thus, it is urgent to reveal their underlying properties by combining the present experimental data with the theoretical analysis. In this work, we extract the structure information of $D_J(3000)^{+,0}$ and $D_J^*(3000)^0$ by performing the phenomenological study, which is an intriguing research topic.

The meson system composed of heavy and light quarks can be described by the heavy quark effective theory, which makes the expansion in terms of $1/m_Q$ suitable when studying the heavy-light meson. Under this framework, the heavy-light meson can be classified by a quantum number $j_\ell^P$, where the angular momentum of a light component $j_\ell$ is a good quantum number in the limit of $m_Q\to \infty$. Thus, the heavy-light mesons can be grouped into different doublets (see Table \ref{Tab: jp} for details).

\tabcolsep=15.5pt
\begin{table}[htbp!]
\renewcommand{\arraystretch}{1.8}
\caption{The doublets categorized by $j_\ell^P$. Here, the subscripts $S$, $P$, $D$, and $F$ denote
the orbital angular momenta $L=0,1,2,$ and $3$, respectively. The superscripts $+$ and $-$ are the corresponding parity $P$. In addition, the quantum numbers in the bracket are $J^P$ with a total angular momentum $J$.
    \label{Tab: jp}}
\begin{tabular}{ccccccc}\toprule[1pt]
$L$&$j_\ell^P$&Doublet&$j_\ell^P$&Doublet \\ \midrule[1pt]
$0$&$\frac{1}{2}^-$&$(0^-,1^-)_S$& -- & --\\
$1$&$\frac{1}{2}^+$&$(0^+,1^+)_P$&
$\frac{3}{2}^+$&$(1^+,2^+)_P$\\
$2$&$\frac{3}{2}^-$&$(1^-,2^-)_D$&
$\frac{5}{2}^-$&$(2^-,3^-)_D$\\
$3$&$\frac{5}{2}^+$&$(2^+,3^+)_F$&
$\frac{7}{2}^+$&$(3^+,4^+)_F$\\
 \bottomrule[1pt]
\end{tabular}
\end{table}

The experimental information on the decays of $D_J(3000)^{+,0}$ and $D_J^*(3000)^0$ can constrain
their $J^P$ quantum numbers. As for $D_J^*(3000)$, its possible $J^P$ quantum number is either $0^+$ or $1^-$ or $2^+$ if $D_J^*(3000)\to D\pi$ occurs via $S$-wave, $P$-wave, and $D$-wave decays, respectively.
As for $D_J(3000)$, its possible $J^P$ quantum number is constrained as either $1^+$ or $0^-/1^-/2^-$ or $1^+/2^+/3^+$ if $D_J(3000)\to D^*\pi$ via $S$-wave, $P$-wave, and $D$-wave, respectively. Although the masses of
$D_J(3000)$ and $D_J^*(3000)$ are close to each other, $D_J(3000)^{+,0}$ and $D_J^*(3000)^0$ are two completely different states {because the angular distributions of $D_J(3000)^{+,0}$ and $D_J^*(3000)^0$
are consistent with unnatural and natural states \cite{Aaij:2013sza}, respectively}. Hence, we can exclude $J^P=1^-,2^+$ assignments to
$D_J(3000)$ and $D_J^*(3000)$ since $D_J(3000)$ and $D_J^*(3000)$ with $J^P=1^-,2^+$ can decay into both $D\pi$ and $D^*\pi$, which is inconsistent with the present experimental data.  By the above analysis, $J^P=0^+$ for $D_J^*(3000)$ and $J^P=1^+,0^-,2^-,3^+$
 for $D_J(3000)$ are possible, which are fully consistent with categorization in terms of natural and unnatural states.
 
{In this work, we examine whether the observed $D_J(3000)$ and $D_J^*(3000)$ states can be grouped into
a conventional charmed meson family. Of course, the observed $D^{*}$ state could be mixed with tetraquark structures and two-meson molecules, for example in
\cite{Vijande:2006hj,Gamermann:2006nm,Gamermann:2007fi} and some like
$D^*\rho$ can also form molecules \cite{Molina:2009eb}. The corresponding studies are also interesting research topics. }

To further extract the structure information on these observed $D_J(3000)^{+,0}$ and $D_J^*(3000)^0$, we will perform the analysis of mass spectrum and the calculation of the corresponding two-body strong decay behaviors.

{\it Spectroscopy}~: As discussed above, the possible quantum numbers for $D_J(3000)^{+,0}$ and $D_J^*(3000)^0$ are deduced by the observed decay channels. Thus, $D_J(3000)^{+,0}$ and $D_J^*(3000)^0$ can be as higher radial excitations and $1F$ states of the $D$ meson family. Before the observations of $D_J(3000)^{+,0}$ and $D_J^*(3000)^0$, different theoretical groups calculated the masses of higher radial excitations of the $D$ meson by different models \cite{Godfrey:1985xj,Matsuki:2007zza,Di Pierro:2001uu,Ebert:2009ua}.

In order to have a better understanding of assignment for newly observed $D_J(3000)$ and $D^*_J(3000)$,
we first discuss the mass spectrum of the $D$ meson, where the relativistic quark model \cite{Godfrey:1985xj} is applied to calculate the masses of higher radial excitations and $1F$ states in the $D$ meson family.
In the original paper \cite{Godfrey:1985xj}, the $D$ meson masses for $1S$, $1P$, $1D$, and $2S$ quantum numbers were given, which are around or below 2.8 GeV. $1S$ and $1P$ states in the $D$ meson family were well established in experiments \cite{Beringer:1900zz}. In addition, there exist several good candidates for $1D$ and $2S$ states in this family since the experimentally observed $D(2550)$, $D(2600)$, and $D(2770)$ can be assigned as $1D$, $2S$ or mixing of $2S$ and $1D$ states \cite{Sun:2010pg}, respectively.
$D_J^*(2650)$, $D_J^*(2760)$, $D_J^*(2580)$, $D_J^*(2740)$ recently reported by LHCb \cite{Aaij:2013sza} can also be identified as the missing $2S$ and $1D$ states when comparing with the former experiment \cite{delAmoSanchez:2010vq} and the predicted $D$ meson mass spectrum.

However, we also notice that the masses of higher radial excitations and $1F$ were not calculated in Ref. \cite{Godfrey:1985xj}, Thus, in our work we calculate the mass of these states in the $D$ meson family with the relativistic quark model, which will be compared with the experimental results of $D_J(3000)$ and $D^*_J(3000)$ discussed here.

We briefly outline calculation in the relativistic quark model below. The total Hamiltonian $\tilde{H}_1$ for the $D$ meson is \cite{Godfrey:1985xj}
\begin{equation}
\tilde{H}_1=\left(p^2+m_1^2\right)^{1/2}+\left(p^2+m_2^2\right)^{1/2}+\tilde{H}_{12}^{\text{conf}}+\tilde{H}_{12}^{\text{so}}
+\tilde{H}_{12}^{\text{hyp}}
\end{equation}
with a confinement term
\begin{equation}
\tilde{H}_{12}^{\text{conf}}=\left(1+\frac{p^2}{E_1E_2}\right)^{1/2}\tilde{G}(r)\left(1+\frac{p^2}{E_1E_2}\right)^{1/2}+\tilde{S}(r).
\end{equation}
Here
$\tilde{H}_{12}^{\text{hyp}}$ is a sum of tensor and contact terms
\begin{equation}
\tilde{H}_{12}^{hyp}=\tilde{H}^{\text{tensor}}_{12}+\tilde{H}^{\text{c}}_{12},
\end{equation}
and $\tilde{H}_{12}^{\text{so}} $ denotes the spin-orbit term, which can  decompose into symmetric $\tilde{H}_{(12)}^{\text{so}}$ and antisymmetric $\tilde{H}_{[12]}^{\text{so}} $. The antisymmetric term  vanishes when the masses of two quarks within the meson are equal. The explicit form of those interactions can found in Appendix A of \cite{Godfrey:1985xj}.

\begin{figure*}[htbp]
\centering%
\begin{tabular}{c}
\scalebox{0.8}{\includegraphics{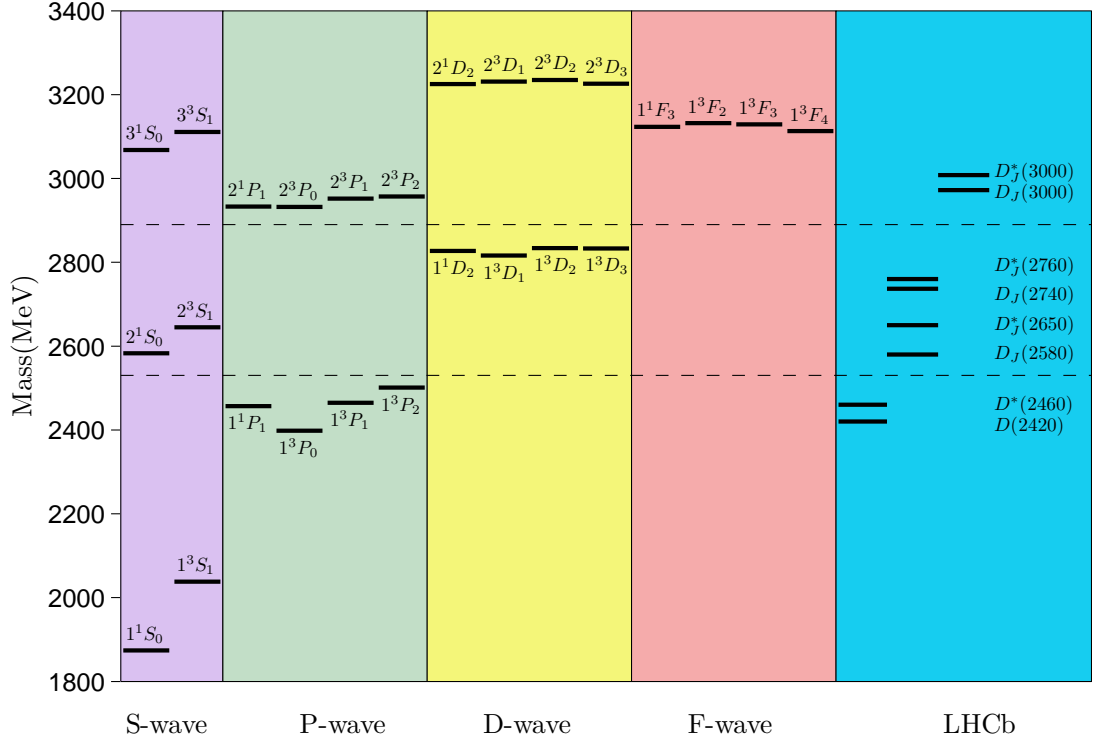}}%
\end{tabular}
\caption{(Color online). The calculated masses of the $D$ meson by the relativistic quark model and the comparison with the experimental values of $D_J(3000)$ and $D^*_J(3000)$. In addition, the information of 
$D_J^*(2760)$, $D_J(2740)$, $D_J^*(2650)$, $D_J(2580)$, $D^*(2460)$ and $D(2420)$ which were reported by LHCb \cite{Aaij:2013sza} is also given here.\label{Fig:mass}}
\end{figure*}

The total Hamiltonian  $\tilde{H}_1$ can be divided into two parts in the bases $|j,m,s,l\rangle$. The first one is the diagonal part $H_{\text{diag}}$, which has the form
\begin{eqnarray}
H_{\text{diag}}&=&\left(p^2+m_1^2\right)^{1/2}+\left(p^2+m_2^2\right)^{1/2}+\tilde{H}_{12}^{\text{conf}}\nonumber\\ &&+\tilde{H}_{(12)}^{\text{so}}+\left(\tilde{H}_{12 }^{\text{tensor}}\right)_{\text{diag}}+\tilde{H}^{\text{c}}_{12},
\end{eqnarray}
and the second one is the off-diagonal part $H_{\text{off}}$
\begin{equation}
H_{\text{off}}=\tilde{H}_{[12]}^{\text{so}}+\left(\tilde{H}^{\text{tensor}}_{12}\right)_{\text{off}},
\end{equation}
where $\left(\tilde{H}_{12}^{\text{tensor}}\right)_{\text{diag}}$ and $\left(\tilde{H}_{12}^{\text{tensor}}\right)_{\text{off}}$ denote the diagonal and off-diagonal parts of $\tilde{H}_{12}^{\text{tensor}}$, respectively.

\renewcommand{\arraystretch}{1.3}
\tabcolsep=3pt
\begin{table}[hbp]\centering
\caption{The comparison of the calculated masses of $D$ mesons from different models \cite{Godfrey:1985xj,Matsuki:2007zza,Di Pierro:2001uu,Ebert:2009ua}. Here, besides reproducing the masses of $1S$, $2S$, $1P$ and $1D$ states in Ref. \cite{Godfrey:1985xj}, we also give the masses of the higher radial excitations and the $1F$ states via the relativistic quark model (see the second column for more details). The symbol "--" denotes that the corresponding masses were not calculated in the corresponding papers. The values marked by "(mixed)" means that mixing of the states with the same $J^P$ quantum number in the heavy quark limit is considered in the calculation}\label{mass}
\begin{tabular}{cccccccc}
\toprule[1pt]
$n^{2S+1}J_L$ & This work  &Ref. \cite{Matsuki:2007zza} & Ref. \cite{Di Pierro:2001uu} &   Ref. \cite{Ebert:2009ua} \\   \midrule[1pt]
$1^1S_0$&1874&1867       &1868                &1871                      \\   
$1^3S_1$&2038&2009       &2005                &2010                      \\   
$2^1S_0$&2583&   --        &2589                &2581                      \\  
$2^3S_1$&2645&  --         &2692                &2632                     \\  
$3^1S_0$&3068&   --        &3141                &3062                        \\ 
$3^3S_1$&3111&   --        &3226                &3096                     \\  
$1^1P_1$&2457&2350 (mixed)&2417 (mixed)         &2469 (mixed)                 \\  
$1^3P_0$&2398&2293       &2377                &2406                    \\  
$1^3P_1$&2465&2432 (mixed)&2490 (mixed)         &2426 (mixed)               \\   
$1^3P_2$&2501&2448       &2460                &2460                    \\   
$2^1P_1$&2933&   --        &2995 (mixed)         &3021                        \\   
$2^3P_0$&2932&    --       &2949                &2919                      \\  
$2^3P_1$&2952&   --        &3045 (mixed)         &2932                        \\   
$2^3P_2$&2957&    --       &3035                &3012                     \\   
$1^1D_2$&2827&    --       &2775 (mixed)         &2850                        \\  
$1^3D_1$&2816&2803       &2795                &2788                        \\   
$1^3D_2$&2834&2726 (mixed)&2833 (mixed)         &2806 (mixed)                 \\  
$1^3D_3$&2833&    --       &2799                &2863                    \\  
$2^1D_2$&3225&   --        &  --                  &3307 (mixed)                        \\   
$2^3D_1$&3231&    --       &   --                 &3228                        \\   
$2^3D_2$&3235&     --      &    --                &3259                        \\ 
$2^3D_3$&3226&     --      &     --               &3335                       \\ 
$1^1F_3$&3123&      --     &      --              &3145 (mixed)                      \\  
$1^3F_2$&3132&    --       &3101                &3090                        \\   
$1^3F_3$&3129&    --       &3123 (mixed)         &3129 (mixed)                     \\  
$1^3F_4$&3113&    --       &3091                &3187                       \\   \bottomrule[1pt]
\end{tabular}
\end{table}

Having the total Hamiltonian, we can obtain the mass spectrum of the $D$ meson family by diagonalizing $H_{\text{diag}}$ in the bases $|j,m,s,l\rangle$ and treating $H_{\text{off}}$ perturbatively. Here, the off-diagonal elements, which can cause mixing, is neglected. 
The calculated mass spectra of the $D$ meson are listed in Table \ref{mass}, where we not only reproduce the results presented in Ref. \cite{Godfrey:1985xj} but also show the values of masses of higher radial excitations and $1F$ states in the $D$ meson family. {In our calculation, we do not include the coupled-channel effects. If considering such effects, the masses of the
observed states can be shifted with respect to the expected $q\bar{q}$
because of the presence of the $D^*\rho,\omega,\phi$ thresholds.}
In addition, we also list the results from other models \cite{Matsuki:2007zza,Di Pierro:2001uu,Ebert:2009ua} for the comparison with those of the relativistic quark model as discussed above. From Table \ref{mass}, we notice that the calculated masses of $D$ mesons from different groups are consistent with each other. What is more important is that the theoretical masses of these $3S$, $2P$ and $1F$ states are around 3 GeV, which are close to the experimental values of $D_J(3000)$ and $D^*_J(3000)$ \cite{Aaij:2013sza}. In Fig. \ref{Fig:mass}, we further compare the results calculated by the relativistic quark model with the experimental results of $D_J(3000)$ and $D^*_J(3000)$, which explicitly shows the above conclusion. This fact indicates that assignment of $D_J(3000)$ and $D^*_J(3000)$ to the candidates of $3S$, $2P$, $2D$ and $1F$ is possible.
However, the analysis of the mass spectrum cannot provide further information of the concrete structure of these two $D_J(3000)$ and $D^*_J(3000)$ states. Thus, we need to carry out their study on two-body strong decay behaviors, which will be useful to further clarify this point.

{\it Decay}~: To describe the OZI allowed two-body strong decays of $D_J(3000)$ and $D^*_J(3000)$ states, the quark pair creation (QPC) model \cite{Micu:1968mk,Le Yaouanc:1972ae,LeYaouanc:1988fx,vanBeveren:1979bd,vanBeveren:1982qb,Bonnaz:2001aj,roberts}  is employed, where the QPC model was extensively applied to study the strong decay of hadrons \cite{Zhang:2006yj,Liu:2009fe,Sun:2009tg,Sun:2010pg,Yu:2011ta,Wang:2012wa,Ye:2012gu,He:2013ttg}.  In the QPC model, meson decay occurs through a quark-antiquark pair created from the vacuum.  Following the
convention in Refs. \cite{Sun:2009tg,Blundell:1996as}, the transition operator $T$ in the non-relativistic limit is expressed as
\begin{eqnarray}
 T&=&-3\gamma\sum_{m}\langle1m;1~-m|00\rangle\int d\textbf{k}_3d\textbf{k}_4\delta^3(\textbf{k}_3+\textbf{k}_4) \nonumber \\
&&\times \mathcal{Y}_{1m}\left(\frac{\textbf{k}_3-\textbf{k}_4}{2}\right)\chi^{34}_{1,-m}
 \phi_0^{34}\omega^{34}_0 d^{\dag}_{3i}(\textbf{k}_3)b^{\dag}_{4j}(\textbf{k}_4),
\end{eqnarray}
which describes a flavor and color singlet quark-antiquark pair created from vacuum with the same quantum number as that of vacuum, namely $J^{PC}=0^{++}$. The flavor and color wave functions have the forms $\phi_0^{34}=(u\bar{u}+d\bar{d}+s\bar{s})/\sqrt{3}$ and $\omega_0^{34}=\delta_{ij}/\sqrt{3}$, respectively. $i$ and $j$ denote the color indices. $\mathcal{Y}_{lm}(\textbf{k})=|\textbf{k}|^lY_{lm}(\textbf{k})$ is the solid harmonic polynomial.  $\chi^{34}_{1,-m}$ is the spin wave function with an angular momentum quantum number $(1,-m)$. $\gamma$ is the model parameter, which describes the strength of quark-antiquark pair creation from vacuum. Here, the $\gamma$ value is chosen to be 6.3 and $6.3/\sqrt{3}$ for the creations of $u/d$ quark and $s$ quark \cite{Sun:2009tg}, respectively.

\renewcommand{\arraystretch}{1.3}
\tabcolsep=2.5pt
\begin{table}[!hbp]\centering
\caption{The adopted masses and $R$ values in  our calculation. Here, the values listed in the fifth column  are from our calculation in the relativistic quark mode.  }\label{wave1}
\begin{tabular}{lcclccccc}
 \\   \toprule[1pt]
States& $R$ (GeV$^{-1}$) \cite{Godfrey:1986wj} &Mass (MeV) \cite{Beringer:1900zz}&States         &$R$   (GeV$^{-1}$)        \\   \midrule[1pt]
 $D^{\pm/0}$           &1.52    &1869.62/1864.86                  & $D(3^1S_0)$          &2.33             \\    
$D_s$                   &1.41    &1968.49                           & $D(3^3S_1)$          &2.38              \\    
$D^{*\pm/0}$         &1.85    &2010.28/2006.98                 & $D(2^1P_1)$            &2.27             \\   
$D_s^*$               &1.69    &2112.3                            &$D(2^3P_0)$            &2.13            \\  
$D(2400)$     &1.85    &2318                             & $D(2^3P_1)$           &2.27             \\    
$D_s(2317)$   &1.75    &2317.8                            &$D(2^3P_2$)          &2.38              \\   
$D(2430)$     &2.00    &2427                              &$D(2^1D_2)$           &2.38             \\    
$D(2420)$     &2.00     &2421.3                           &$D(2^3D_1$)             &2.27           \\    
$D(2460)$     &2.22      &2464.4                          &$D(2^3D_2)$           &2.38             \\    
$K^{\pm/0}$           &1.41      &493.68 /497.61                &$D(2^3D_3) $            &2.56               \\  
$K^*$         &2.08     &891.66                           & $D(1^1F_3)$           &2.44            \\   
$\pi^{\pm/0}$         &1.41      &139.57/134.98                   & $D(1^3F_2)$          &2.33             \\   
$\eta$        &1.41      &547.85                          &$D(1^3F_3)$          &2.38              \\  
$\eta'$       &1.41      &957.78                        &$D(1^3F_4)$            &2.56              \\  
$\rho$        &2.08      &775.49     &--&--                                                              \\   
 $\omega$      &2.08     &782.65 &      --&--                                          \\   \bottomrule[1pt]
\end{tabular}
\end{table}

The  helicity amplitude $\mathcal{M}^{M_{J_A}M_{J_B}M_{J_C}}$(\textbf{K}) for a decay of $A$ meson into $B+C$ is defined as
\begin{equation}
\langle BC|T|A\rangle=\delta^3(\textbf{K}_B+\textbf{K}_C-\textbf{K}_A)\mathcal{M}^{M_{J_A}M_{J_B}M_{J_C}}(\textbf{K}),
\end{equation}
where $|A\rangle$, $|B\rangle$ and $|C\rangle$ denote mock states \cite{Hayne:1981zy}. The expression of a mock state for a meson $A$, for example, is given by
\begin{eqnarray}
|A(n^{2S+1}L_{JM_J})(\textbf{K}_A)\rangle&=&\sqrt{2E}\sum_{M_S,M_L}\langle LM_LSM_S| JM_J\rangle \chi^{A}_{SM_S} \nonumber \\
&&\times\phi^{A}\omega^{A}\int d\textbf{k}_1d\textbf{k}_2\delta^3(\textbf{K}_A-\textbf{k}_1-\textbf{k}_2)
\nonumber \\
&&\times\Psi_{nLM_L}^A(\textbf{k}_1,\textbf{k}_2)|q_1(\textbf{k}_1)
\bar{q}_2(\textbf{k}_2)\rangle,
\end{eqnarray}
where 
$\chi^A_{SM_S}$, $\phi^A$ and $\omega^A$ denote the spin, flavor and color wave functions of a meson $A$, respectively. $\Psi^A_{nLM_L}(\textbf{k}_1,\textbf{k}_2)$ is the meson spacial wave function in momentum space with the form
\begin{equation}
\Psi_{nLM_L}^A(\textbf{k}_1,\textbf{k}_2) =R^A_{nL}(|\textbf{k}|)Y_{LM_L}(\textbf{k}),
\end{equation}
where $R^A_{nL}(|\textbf{k}|)$ is the radial wave function and $\textbf{k}=(m_1\textbf{k}_2-m_2\textbf{k}_1)/(m_1+m_2)$ is the relative momentum between quark and antiquark. In our calculation, the simple harmonic oscillator (SHO) wave function is chosen to represent the radial wave function. Thus, the parameter $R$ in the SHO wave function is involved in our calculation,
where the $R$ values in the SHO wave functions can be determined by reproducing the root mean square (RMS) of the corresponding states calculated in the relativistic quark model. In Table \ref{wave1}, we list these adopted $R$ values in our calculation of strong decays.

The corresponding partial wave amplitude $\mathcal{M}_{SL}(|\textbf{K}|)$ is related to the helicity amplitude $\mathcal{M}^{M_{J_A}M_{J_B}M_{J_C}}(\textbf{K})$ via the Jacob-Wick formula \cite{Jacob:1959at} by choosing $\textbf{K}$ along the positive $z$ axis, i.e.,
\begin{eqnarray}
\mathcal{M}_{LS}(\textbf{K})&=&\frac{\sqrt{2L+1}}{2J_A+1}\sum_{M_{J_B},M_{J_C}}\langle L0SM_{J_A}|J_AM_{J_A}\rangle \nonumber \\
 &&\times\langle J_BM_{J_B}J_CM_{J_C}|SM_{J_A}\rangle
\mathcal{M}^{M_{J_A}M_{J_B}M_{J_C}}(\textbf{K}).
\end{eqnarray}
The two body decay width in terms of the partial wave amplitude is
\begin{equation}
\Gamma=\pi^2\frac{|\textbf{K}|}{M_A^2}\sum_{LS}|\mathcal{M}_{LS}|^2
\end{equation}
with $M_A$ is the mass of a particle $A$.

We need to emphasize that the mixture between $2^1P_1$ and $2^3P_1$ is considered, i.e.,
\begin{equation}
\left(
  \begin{array}{c}
    2P(1^+) \\
    2P^\prime(1^+)  \\
  \end{array}
\right) =
\left(
\begin{array}{cc}
   \cos\theta_1&\sin\theta_1 \\
   -\sin\theta_1&\cos\theta_1  \\
\end{array}
\right)
\left(
  \begin{array}{c}
    2^1P_1 \\
    2^3P_1 \\
  \end{array}
\right),\label{h1}
\end{equation}
where $\theta_1$ is the mixing angle. Later, we will discuss it by taking the phenomenological analysis into account.  In addition, we also introduce the mixtures of $2^1D_2$ and $2^3D_2$ states and $1^1F_1$ and $1^3F_3$ states, which satisfy
\begin{equation}
\left(
  \begin{array}{c}
    2D(1^-) \\
    2D^\prime(1^-)  \\
  \end{array}
\right) =
\left(
\begin{array}{cc}
    \cos\theta_2&\sin\theta_2 \\
    -\sin\theta_2&\cos\theta_2  \\
\end{array}
\right)
\left(
  \begin{array}{c}
    2^1D_2 \\
    2^3D_2 \\
  \end{array}
\right)\label{h2}
\end{equation}
and
\begin{equation}
\left(
  \begin{array}{c}
    1F(3^+) \\
    1F^\prime(3^+)  \\
  \end{array}
\right) =
\left(
\begin{array}{cc}
    \cos\theta_3&\sin\theta_3 \\
    -\sin\theta_3&\cos\theta_3  \\
\end{array}
\right)
\left(
  \begin{array}{c}
    1^1F_3 \\
    1^3F_3 \\
  \end{array}
\right),\label{h3}
\end{equation}
respectively. Here, $\theta_2$ and $\theta_3$ are the corresponding mixing angles. In Eqs. (\ref{h1})-(\ref{h3}), we use prime to distinguish  the states with the same $J^P$ quantum number.

In our calculation, the final states are related to $D(2420)/D(2430)$ and $D_s(2460)/D_s(2536)$, which
are the $1^+$ states in the $D$ and $D_s$ meson families, respectively.  $D(2420)/D(2430)$ and $D_s(2460)/D_s(2536)$ are the mixing of $1^1P_1$ and $1^3P_1$ states \cite{Barnes:2005pb, Matsuki:2010zy}, i.e.,
\begin{equation}
\left(
  \begin{array}{c}
    D(2430) \\
    D(2420)  \\
  \end{array}
\right) =
\left(
\begin{array}{cc}
   \cos\theta&\sin\theta \\
   -\sin\theta&\cos\theta  \\
\end{array}
\right)
\left(
  \begin{array}{c}
    D(1^1P_1) \\
    D(1^3P_1) \\
  \end{array}
\right)
\end{equation}
and
\begin{equation}
\left(
  \begin{array}{c}
    D_s(2460) \\
    D_s(2536)  \\
  \end{array}
\right) =
\left(
\begin{array}{cc}
   \cos\theta&\sin\theta \\
   -\sin\theta&\cos\theta  \\
\end{array}
\right)
\left(
  \begin{array}{c}
    D_s(1^1P_1) \\
    D_s(1^3P_1) \\
  \end{array}
\right),
\end{equation}
where the mixing angle is taken as $\theta=-54.7^\circ$ or $35.3^\circ$, which is the estimate in the heavy quark limit.

{\it Phenomenological analysis}~;
{\it Case $D_{J}^*(3000)$}~:
  As discussed above, $D_{J}^*(3000)$ can be a natural parity state. Thus, we study its decay behavior with the $3^3S_1$, $2^3D_1$, $2^3D_3$, $2^3P_0$, $2^3P_2$, $1^3F_2$, and $1^3F_4$ assignments. The partial and total decay widths of $D_J^*(3000)$ are shown in Table \ref{width1}. Our results show that the $3^3S_1$, $2^3D_1$, $2^3D_3$, $2^3P_2$, and $1^3F_4$ assignments to $D_J^*(3000)$ can be fully excluded since the corresponding total decay widths are far less than the full width of $D_J^*(3000)$ measured by LHCb \cite{Aaij:2013sza}.  The remaining two possibilities of the internal structure of $D_J^*(3000)$ are $2^3P_0$ and $1^3F_2$, where the total decay width of these states have the same oder of magnitude as the experimental result.

If $D_J^*(3000)$ is a  $2^3P_0$ state, the $D\pi$ channel is dominant while the $D^*\pi$ channel is fully forbidden, which is consistent with the experimental observation. In addition,  $D^*\rho$, $D(2420\pi)$, and $D(2427)\pi$ are the dominant decay channels of $D_J^*(3000)$ and the main decay channels of  $D_J^*(3000)$ include $D\eta$, $D_sK$, and $D^*\omega$.

If $D_J^*(3000)$ is a $2^3F_2$ state, $D_J^*(3000)$ dominantly decays into $D(2427)\pi$. $D^*\pi$, $D\pi$,
$D\rho$, $D^*\rho$, $D(2460)\pi$, and $D(2420)\pi$ are the main decay modes. Among these main decay modes, the partial decay widths of $D_J^*(3000)$ into $D\pi$ and $D^*\pi$ are comparable with each other, which indicates that $D_J^*(3000)$ can be found in the $D^*\pi$ channel. However, there is no evidence of $D_J^*(3000)$ in the $D^*\pi$ invariant mass spectrum given by LHCb \cite{Aaij:2013sza}. According to this fact, we can exclude the $2^3F_2$ assignment to $D_J^*(3000)$.

Although the $3^3S_1$, $2^3D_1$, $2^3D_3$, $2^3P_2$, $2^3F_2$, and $1^3F_4$ assignments to $D_{J}^*(3000)$ can be excluded by the above calculation, we obtain the decay behaviors of these states, where we set the masses of these states to be the mass of $D_J^*(3000)$. The numerical results in Table \ref{width1} show:
\begin{enumerate}
\item The $3^3S_1$ state mainly decays into $D^*\pi$, $D^*\rho$, $D(2460)\pi$, $D(2420)\pi$, and $D(2427)\pi$. Its
$D_sK$, $D^*\eta^\prime$, $D_sK^*$, $D_s^*K^*$, $D(2420)\eta$, $D(2427)\eta$ and $D_s(2460)K$ are tiny. In general, the  $3^3S_1$ state is a narrow $D$ meson.
\item As for the $2^3D_1$ state, the main decay channels include $D\pi$, $D\eta$, $D_sK$, $D^*\pi$, $D^*\rho$, $D^*\omega$, and $D(2427)\pi$. Among main decay modes, $D\pi$ is dominant. Hence, the ideal decay channel to experimentally search for the $2^3D_1$ state is $D\pi$.
\item The $2^3D_1$ state also has a very narrow width. $D\pi$, $D^*\rho$, $D^*\omega$, $D(2460)\pi$, and $D(2420)\pi$ are its main decay modes.
\item The $2^3P_2$ state can mainly decay into $D(2420)\pi$, $D(2460)\pi$, $D\rho$, $D(2427)\eta$ and $D\pi$. The sum of its partial decay widths can reach up to 47 MeV.
\item For the $1^3F_4$ state, $D^*\rho$ is a dominant decay mode while $D^*\omega$ is another main decay channel. The total decay width of $1^3F_4$ state is 39 MeV.
\end{enumerate}
The above information is valuable to further search for the partners of $D_J^*(3000)$.

{\it Case $D_J(3000)$} :
In the following, we discuss $D_J(3000)$ by combining our results with the experimental data. As for $D_J(3000)$, the possible quantum numbers include $0^-$, $1^+$, $2^-$ and $3^+$, which correspond to the $3^1S_0$, $2P(1^+)/2P^\prime(1^+)$, $2D(2^-)/2D^\prime(2^-)$ and $1F(3^+)/1F^\prime(3^+)$, respectively. In Table \ref{width2}, the partial and total decay widths of $D_J(3000)$ as the $3^1S_0$ state are shown and the allowed decay channels of  $D_J(3000)$ as the $2P(1^+)/2P^\prime(1^+)$, $2D(2^-)/2D^\prime(2^-)$ and $1F(3^+)/1F^\prime(3^+)$ states are given. Since the total width of $D_J(3000)$ as the $3^1S_0$  state is about 25 MeV, which is far smaller than the experimental width. Hence, the $3^1S_0$ assignment to $D_J(3000)$ can be fully excluded.

\begin{figure}[!htbp]
\centering%
\begin{tabular}{c}
\scalebox{0.9}{\includegraphics{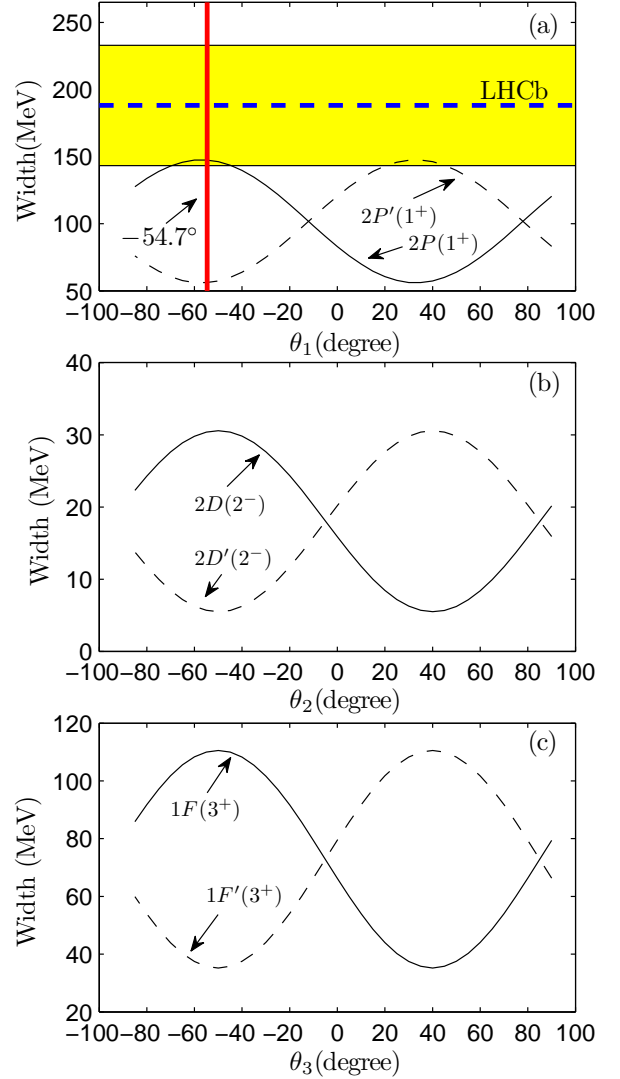}}%
\end{tabular}
\caption{(Color online). The dependences of the total decay width of $D_J(3000)$ as the
$2P(1^+)/2P^\prime(1^+)$, $2D(2^-)/2D^\prime(2^-)$ and $1F(3^+)/1F^\prime(3^+)$ states on
the mixing angles $\theta_1$, $\theta_2$ and $\theta_3$ respectively. Here, the blue dashed line with a yellow band is the experimental width of $D_J(3000)$ by LHCb \cite{Aaij:2013sza}. The red vertical solid line corresponds to the mixing angle $\theta_1=-54.7^\circ$. \label{Fig:mass11}}
\end{figure}

In Fig. \ref{Fig:mass11}, we present the total decay width of  $D_J(3000)$ as the
$2P(1^+)/2P^\prime(1^+)$, $2D(2^-)/2D^\prime(2^-)$ and $1F(3^+)/1F^\prime(3^+)$ states, which are dependent on the mixing angles $\theta_1$, $\theta_2$ and $\theta_3$, respectively.  The Fig. \ref{Fig:mass11} (a) shows that the total decay width of the $2P(1^+)$ state overlaps with the experimental result with an error of $D_J(3000)$ when taking $\theta_1=(-40\sim-70)^\circ$ or $\theta_1=(20\sim45)^\circ$.
Here, both regions are consistent with the requirement from the heavy quark limit since
$\theta_1=-54.7^\circ$ or $\theta_1=35.3^\circ$ can be estimated in the heavy quark limit \cite{Godfrey:1986wj,{Matsuki:2010zy}}. The difference is which state is identified as $P$ or $P'$.
Hence, explanation of $D_J(3000)$ as the $2P(1^+)$ state in the $D$ meson family is suitable. As the partner of
this state, the $2P'(1^+)$ state has narrow total decay width, which can be grouped into the $1^+$ state in the $(1^+,2^+)_P$ doublet. $D_J(3000)$ is the $1^+$ state in the $(1^+,2^+)_P$ doublet since $D_J(3000)$ is of broad width. These conclusions are also in good agreement with the estimate by the heavy quark limit.

Since the total decay widths of $D_J(3000)$ as the $2D(2^-)/2D^\prime(2^-)$ and $1F(3^+)/1F^\prime(3^+)$ states are deviated from the experimental width of $D_J(3000)$ (see Fig. \ref{Fig:mass11} (c)-(d) for more details), we exclude these two assignments to $D_J(3000)$.

Besides discussing the information of the total decay width, we further give the partial decay widths of $D_J(3000)$ as the $2P(1^+)$ state, which is shown in Table \ref{width2}. In addition, the partial decay widths of the $2P^\prime(1^+)$ are also calculated here.

From Table \ref{width2}, we notice that the $D^*\pi$ channel is one of the most dominant decay modes
of the $2P(1^+)$, which is well consistent with the experimental observation \cite{Aaij:2013sza}, where $D_J(3000)$ was first observed in the $D^*\pi$ decay channel. Other dominant decay modes of the $2P(1^+)$ state include $D(2460)\pi$, $D^*\rho$, $D(2420)\pi$ and $D(2427)\pi$, while the main decays are $D^*\eta$, $D_s^*K$, $D\rho$, $D\omega$, $D^*\omega$, and $D(2400)\pi$.
As for the $2P^\prime(1^+)$ state, it can dominantly decay into $D^*\rho$, $D(2400)\pi$, $D(2420)\pi$ and $D(2427)\pi$, which are valuable to search for the $2P^\prime(1^+)$ state in experiment.

\renewcommand{\arraystretch}{1.4}
\tabcolsep=12.5pt
 \begin{table*}[htbp]\centering
 \caption{The obtained partial and total decay widths of $D^*_J(3000)$ with several possible assignments. If the corresponding decay channel is forbidden, we mark it by "--". All values are in units of MeV. }\label{width1}
\begin{tabular}{lcccccccccccc}
 \\   \toprule[1pt]
Channels       &$3^3S_1$           &$2^3D_1$            &$2^3D_3$           &$2^3P_0$&$2^3P_2$             &$1^3F_2$            &$1^3F_4$
\\   \midrule[1pt]
$D\pi$         &0.91               &18                  & 1.3               &49      &1.8                  & 16                & 1.2
 \\   
$D\eta$        &0.25               &2.7                 & 0.11              &8.8     &0.11                 & 2.6                & 0.77
  \\   
$D\eta'$       &0.13               &0.59                &$2.8\times 10^{-2}$&2.7     &$2.9\times 10^{-4}$  & 0.58      & $2.3\times 10^{-3}$
 \\   
$D_sK$         &$9.6\times 10^{-2}$&1.6                 &$4.9\times 10^{-2}$&6.6     &0.13                 & 1.1       & $1.8\times 10^{-2}$
 \\  
$D^*\pi$       &3.5                &4.5                 &0.21               &--      &$8.1\times 10^{-3}$  & 13                 & 1.8
 \\   
$D^*\eta$      &0.51               &0.41                &$6.7\times 10^{-3}$&--      &$9.1\times 10^{-2}$  & 1.8                & 0.11
 \\   
$D^*\eta'$      &$5.3\times10^{-3}$ &$1.3\times 10^{-4}$ &$1.2\times 10^{-5}$&--     &$8.0\times 10^{-2}$  &$2.6\times 10^{-2}$ & $1.7\times 10^{-5}$ \\   
$D_s^*K$       &0.24               & 0.22               &$9.5\times 10^{-3}$&--      &$4.5\times 10^{-3}$  & 0.67               & $1.8\times 10^{-2}$    \\   
$D\rho$        &0.42               &$1.1\times 10^{-3}$ &$6.3\times 10^{-2}$&--      &3.2                  & 8.1                & 0.54
\\   
$D\omega$      &0.13               &$6.7\times 10^{-4}$ &$2.1\times 10^{-2}$&--      &1.1                  & 2.6                & 0.18
  \\   
$D_sK^*$       &$6.6\times10^{-3}$ &$4.3\times 10^{-2}$ &$3.5\times 10^{-4}$&--      &$8.5\times 10^{-2}$  &$7.9\times 10^{-2}$ & $7.7\times 10^{-4}$  \\  
$D^*\rho$      &1.7                & 3.7                & 0.52              &41      &10                   & 5.1                & 25.0
 \\  
$D^*\omega$    &0.61               & 1.2                & 0.16              &13      &3.1                  & 1.6                & 8.1
 \\   
$D_s^*K^*$     &$6.6\times10^{-3}$ &$1.6\times 10^{-3}$ &$7.4\times 10^{-4}$&1.0     &1.1                  &$2.8\times 10^{-5}$ &$1.7\times 10^{-4}$
   \\   
$D(2460)\pi$   &2.4                & 2.5                & 0.43              &--      &4.7                  & 8.3                &1.0
 \\   
$D(2420)\pi$   &6.2                & 4.5                & 1.5               &38      &18                   & 8.4                & 0.92
   \\   
$D(2420)\eta$  &$3.1\times10^{-3}$ &$5.3\times 10^{-2}$ &$5.1\times 10^{-3}$&1.1     &0.51                 & 0.16               &$1.8\times 10^{-4}$   \\   
$D(2427)\pi$   &1.4                & 4.9                &0.12               &30      &2.7                  & 62                & $3.8\times 10^{-2}$   \\   
$D(2427)\eta$  &$4.5\times10^{-3}$ & 0.32               &$1.6\times 10^{-4}$&0.91    &$1.5\times 10^{-2}$  & 1.3                & $2.6\times10^{-6}$  \\   
$D_s(2460)K$   &$6.1\times10^{-3}$ & 0.11               &$2.5\times 10^{-4}$&1.5     &$2.2\times 10^{-2}$  & 1.8                & $5.5\times10^{-6}$  \\   \midrule[1pt]
Total width    &18                 & 45                 &5                  &194     & 47                  &136                 &  39
 \\   \bottomrule[1pt]
\end{tabular}
\end{table*}

\tabcolsep=20pt
 \begin{table*}[!hbp]\centering
 \caption{The partial and total decay widths (in units of MeV) of $D_J(3000)$ as the $D$ meson with the $3^1S_0$,  $2P(1^+)$ and $2P^\prime(1^+)$ quantum numbers. The partial widths of  the $2D(2^-)/2D^\prime(2^-)$ and $1F(3^+)/1F^\prime(3^+)$  states are not given since we can exclude these possibilities after discussing their partial decay widths that depend  on the corresponding mixing angle. Here, the forbidden and allowed decay channels are marked with symbols  "--" and $\square$, respectively. To present the results of the $2P(1^+)$ and $2P^\prime(1^+)$ states, we fix the corresponding mixing angle as $\theta_1=-54.7^\circ$.  }\label{width2}
\begin{tabular}{lcccccccccccc}
 \\   \toprule[1pt]
Channels        &$3^1S_0$          &$2P(1^+)$&$2P^\prime(1^+)$   &$2D(2^-)/2D^\prime(2^-)$&$1F(3^+)/1F^\prime(3^+)$           \\   \midrule[1pt]
$D^*\pi$        &4.8               &38                 &1.3                &$\square$&$\square$   \\  
$D^*\eta$       &0.52              &5.2                &0.49               &$\square$&$\square$  \\  
$D^*\eta'$      &$2.5\times10^{-6}$&$2.3\times 10^{-2}$&$2.6\times 10^{-4}$&$\square$&$\square$   \\ 
$D_s^*K$        &0.26              &3.7                &$9.9\times 10^{-2}$&$\square$& $\square$  \\   
$D\rho$         &$5.9\times10^{-2}$&7.6                &4.7                &$\square$&$\square$   \\  
$D\omega$       &$1.7\times10^{-2}$&2.5                &1.5                &$\square$&$\square$    \\   
$D_sK^*$        &$4.1\times10^{-6}$&0.12               &0.7                &$\square$&$\square$   \\   
$D^*\rho$       &4.3               &15                 &14               &$\square$&$\square$  \\   
$D^*\omega$     &1.5               &4.9                &4.6                &$\square$&$\square$  \\   
$D(2400)\pi$    &12                &6.0                &11                &$\square$&$\square$  \\   
$D(2400)\eta$   &0.26              &$6.8\times 10^{-2}$&0.14               &$\square$&$\square$  \\  
$D_s(2317)K$    &0.85              &0.67               &1.2                &$\square$&$\square$  \\   
$D(2460)\pi$    &5.1               &38                 &3.3                &$\square$&$\square$  \\  
$D(2420)\pi$    &--                &14                 &8.8                &$\square$&$\square$  \\   
$D(2420)\eta$   &--                &$4.2\times 10^{-3}$&$2.3\times 10^{-3}$&$\square$&$\square$   \\  
$D(2427)\pi$    &--                &11                 &5.3                &$\square$&$\square$   \\   
$D_s(2460)K$    &--                &$8.2\times 10^{-2}$&$4.5\times 10^{-2}$&$\square$&$\square$  \\  \midrule[1pt]
Total width     &30                &147                &56 &      &                 \\   \bottomrule[1pt]
\end{tabular}
\end{table*}

{\it Summary}~: Stimulated by two newly observed $D_J(3000)$ and $D_J^*(3000)$ \cite{Aaij:2013sza}, we carry out the study of their properties by analyzing the mass spectrum and calculating the decay behaviors. This phenomenological investigation can further shed light on the underlying structures of $D_J(3000)$ and $D_J^*(3000)$.

Both the obtained mass spectra of the $D$ meson and the two-body decay behaviors calculated by the QPC model support the $2P(1^+)$ and $2^3P_0$ assignments to $D_J(3000)$ and $D_J^*(3000)$, respectively, {since the widths of other states are very small or they can decay into both $D\pi$ and $D^*\pi$. We note that $D_J(3000)$ and $D^{*}_J(3000)$ states are strongly correlated to the background
parameters as shown in Ref. \cite{Aaij:2013sza}. Thus, more experimental data are needed to
state a clear conclusion on the existence of $D_J(3000)$ and $D^{*}_J(3000)$ states,
their properties and further identification of the obtained states
in Ref. \cite{Aaij:2013sza}.}
In our work, the information of other decay modes of $D_J(3000)$ and $D_J^*(3000)$ is given with the above quantum number assignments, which is important to further experimentally study these two observed $D_J(3000)$ and $D_J^*(3000)$ resonances.

When investigating $D_J(3000)$ and $D_J^*(3000)$, we have also given the decay behaviors of the $3^1S_0$, $3^3S_1$,  $2D(2^-)$, $2^3D_1$, $2D^\prime(2^-)$, $2^3D_3$,
$2P^\prime(1^+)$, $2^3P_2$,  $1F(3^+)$, $1^3F_2$, $1F^\prime(3^+)$, and $1^3F_4$ states, which are still missing in the  $D$ meson family. As the accumulation of more experimental data, more $D$ mesons will be found in future experiments. Thus, our numerical results can provide valuable hints to experimental study on these  missing $3^1S_0$, $3^3S_1$,  $2D(2^-)$, $2^3D_1$, $2D^\prime(2^-)$, $2^3D_3$,
$2P^\prime(1^+)$, $2^3P_2$,  $1F(3^+)$, $1^3F_2$, $1F^\prime(3^+)$, and $1^3F_4$ in the $D$ meson family.

At the present, there are good candidates for these low lying states in the $D$ meson family \cite{Beringer:1900zz}. The experimental and theoretical study of higher radial excitations and $1F$ states in the $D$ meson family is interesting and important research field. In the past years, the BaBar \cite{delAmoSanchez:2010vq} and LHCb \cite{Aaij:2013sza} experiments have made some progress on this field. We also expect more experimental efforts will be done in future. As a research filed full of challenges and chances, further theoretical study on higher radial excitations and $1F$ states in the $D$ meson family is also necessary utilizing different approaches and models.

{Note added: After completion of this work, a theoretical paper on $D_J(2580)$, $D_J^*(2650)$, $D_J(2740)$, $D^*_J(2760)$, $D_J(3000)$, and $D_J^*(3000)$ have appeared \cite{Wang:2013tka}. Using the heavy meson effective theory, their strong partial decays, especially the corresponding ratios, are studied. 
}

\vfil
\section*{Acknowledgements}

This project is supported by the National Natural Science
Foundation of China under Grants No. 11222547, No. 11175073 and No. 11035006, the Ministry of Education of China
(FANEDD under Grant No. 200924, SRFDP under Grant No.
2012021111000, and NCET), the Fok Ying Tung Education Foundation
(No. 131006).

\end{document}